\documentclass[conference]{IEEEtran}
\usepackage[margin=0.7in]{geometry}

\usepackage[T1]{fontenc}

\usepackage{cite}
\usepackage{graphicx}
\graphicspath{{Figure/}}
\usepackage[cmex10]{amsmath}
\usepackage{amsthm}
\usepackage{amssymb}
\usepackage{algorithmic}
\usepackage{array}
\usepackage{color}
\usepackage{epstopdf}
\usepackage{amsfonts}

\usepackage{pgfplots}
\pgfplotsset{compat=1.3}
\usepackage{tikz}							
\usetikzlibrary{shapes}
\usetikzlibrary{spy}
\usetikzlibrary{circuits}
\usetikzlibrary{arrows}
\usepackage{subfig}

\newcommand{\mat}[1]{\boldsymbol{\mathrm{#1}}}

\newcommand{\tr}{\mathrm{tr}}

\usepackage{mathtools}

\begin{document}
\linespread{1}
\title{Impact of Realistic Propagation Conditions\\ on Reciprocity-Based Secret-Key Capacity}

\author{\IEEEauthorblockN{François Rottenberg\IEEEauthorrefmark{1}\IEEEauthorrefmark{2},
		Philippe De Doncker\IEEEauthorrefmark{2},
		François Horlin\IEEEauthorrefmark{2} and
		Jérôme Louveaux\IEEEauthorrefmark{1}
}
	\IEEEauthorblockA{\IEEEauthorrefmark{1}ICTEAM institute, Université catholique de Louvain, Belgium
	}
	\IEEEauthorblockA{\IEEEauthorrefmark{2}OPERA department, Université libre de Bruxelles, Belgium
	}
}

\maketitle

\begin{abstract}
Secret-key generation exploiting the channel reciprocity between two legitimate parties is an interesting alternative solution to cryptographic primitives for key distribution in wireless systems as it does not rely on an access infrastructure and provides information-theoretic security. The large majority of works in the literature generally assumes that the eavesdropper gets no side information about the key from her observations provided that (i) it is spaced more than a wavelength away from a legitimate party and (ii) the channel is rich enough in scattering. In this paper, we show that this condition is not always verified in practice and we analyze the secret-key capacity under realistic propagation conditions.

\end{abstract}


\section{Introduction}\label{section:Introduction}


The secrecy-capacity is defined as the number of bits per channel use that can be reliably transmitted to a legitimate receiver (Bob) while guaranteeing a negligible information leakage to the eavesdropper (Eve). The seminal work of Wyner \cite{Wyner1975} and its extension to more general channels \cite{Csiszar1978} have shown that a "physical advantage" at Bob with respect to Eve is required to guarantee a larger-than-zero secrecy capacity. This "physical advantage" implies that Eve channel has to be noisier, which might not be always verified in practice \cite{bloch2011physical}. 

\begin{figure}[!t]  
	\centering
	
	\resizebox{0.85\textwidth}{!}{%
		{\includegraphics[clip, trim=0cm 14cm 12cm 0cm, scale=1]{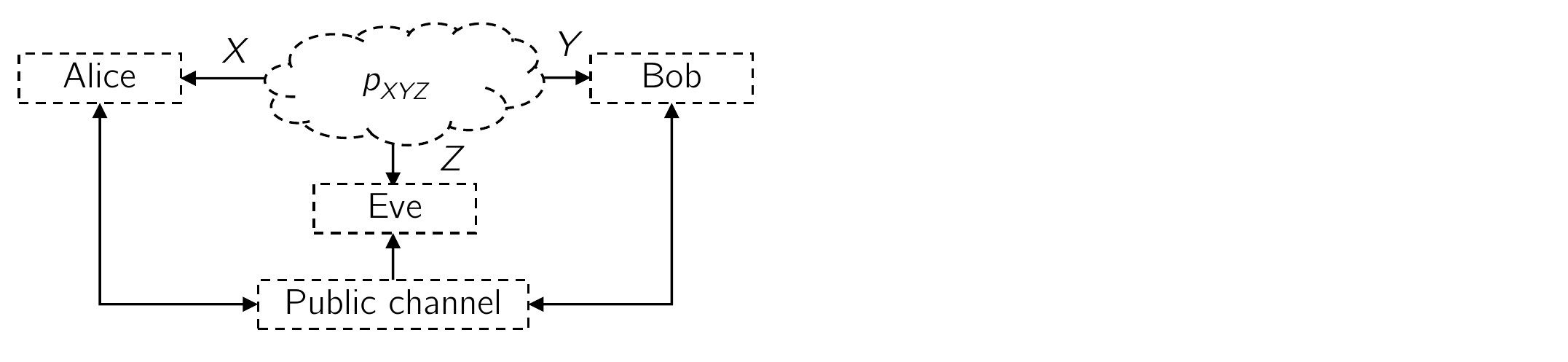}} 
	}
	\vspace{-1em}
	\caption{Source model for secret-key agreement.}
	\label{fig:source_model}
	\vspace{-2em}
\end{figure}

Later, the works of \cite{Maurer1993,Csiszar1993} have shown that the pessimistic limitation of requiring an advantage over the eavesdropper can be overcome by using stronger communication schemes that are not restricted to one-way rate-limited communications \cite{bloch2011physical}. Maurer \cite{Maurer1993} and Ahlswede and Csisz\'{a}r \cite{Csiszar1993} were the first to analyze the problem of generating a secret key from correlated observations. In the source model (see Fig.~\ref{fig:source_model}), two legitimate parties (Alice and Bob) and one illegitimate party (Eve) observe the realizations of a discrete memoryless source. From their observations, Alice and Bob have to distill an identical key that remains secret from Eve. Moreover, Alice and Bob have access to a public error-free authenticated channel with unlimited capacity. This helps them to agree a common key with the limitation that Eve also can listen to it. Upper and lower bounds for the secret-key capacity, defined as the number of secret bits that can be generated per observation of the source, were derived in \cite{Csiszar1993,Maurer1993}. Their results show that positive secret-key rates are achievable even if the channel from Alice to Bob is not degraded with respect to the channel from Alice to Eve. 

A practical source of common randomness at Alice and Bob consists of the wireless channel reciprocity, which implies that the propagation channel from Alice to Bob and from Bob to Alice is identical if both are measured within the same channel coherence time and at the same frequency. At each coherence time, Alice and Bob can repeatedly sample the channel by sending each other a pilot symbol so as to obtain a set of $N$ highly correlated observations and finally start a key-distillation procedure. The vast majority of works in the literature considers that Eve gets no side information about the key from her observations, which consist of the pilots transmitted by Alice and Bob \cite{AzimiSadjadi2007,wong2009secret,Ye2010,Chen2011,Jorswieck2013}. Often, this assumption is justified by the fact that: (i) Eve is supposed to be separated from Bob and Alice by more than one wavelength (otherwise she could be easily detected) and (ii) the channel environment is supposed to be rich enough in scattering implying that the fading process of the channels at the different antennas can be considered independent. The assumption of rapid decorrelation in space has been further validated through measurement campaigns \cite{AzimiSadjadi2007,Mathur2008,Madiseh2009,Ye2010,Zhang2016}. Moreover, this assumption is convenient as it drastically simplifies the problem of secret-key generation. Indeed, the secret-key capacity simply becomes equal to the mutual information between Alice and Bob since Eve cannot learn anything about the key from her channel observations. This implies that Eve can only learn about the secret-key from the discussion over the public channel.  

However, it often occurs in practical scenarios that scatterers are clustered with small angular spread rather than being uniformly distributed, which leads to much longer spatial decorrelation length. Quite surprisingly, only a few works have considered the impact of potential spatial correlation at Eve's side. For instance, ref. \cite{Chou2010} studied the impact of channel sparsity, inducing correlated eavesdropping, on secret-key capacity. However, no advanced physical characterization of the propagation environment was considered. In \cite{Zhang2017}, correlation of the eavesdropper channel is taken into account but the spatial and time correlation of the channel is modeled according to Jakes Doppler model, which again assumes a rich scattering environment and leads to decorrelation in space after only a few wavelengths.

In the light of these limitations, we feel that a deep theoretical analysis of the impact of realistic propagation environments on the secret-key capacity is still missing in the literature. In this paper, as opposed to most of the works in the literature, we characterize the secrecy-key capacity without making the simplifying assumption of a rich scattering environment. The obtained lower and upper bounds for the secrecy-key capacity takes into account the impact of realistic propagation environments, which controls the spatial decorrelation of Eve. Eve is also allowed to have a more powerful receiver than Alice and Bob, resulting in a larger signal-to-noise ratio (SNR). We show that, under typical propagation conditions drawn from 3GPP models \cite{3GPP_TR_38_901_v15}, current evaluations of the secret-key capacity, relying on a rich scattering assumption, are too optimistic as they underestimate the information available at Eve's side. New 5G technologies strongly relying on directive transmission, such as Massive MIMO and milimeter-wave communications, are other typical examples of small angular spreads that can lead to long spatial decorrelation lengths.





\textbf{Notations}: 
Vectors and matrices are denoted by bold lowercase and uppercase letters, respectively (resp.). Non bold upper case letter refers to a random variable. The vector notation $\vec{r}$ refers to a 3D position. Superscripts $^*$, $^T$ and $^H$ stand for conjugate, transpose and Hermitian transpose operators. The symbols $\tr$, $\mathbb{E}$, $\Im$ and $\Re$ denote the trace, expectation, imaginary and real parts, respectively. $\jmath$ is the imaginary unit. The norm $\|\mat{A}\|$ is the Frobenius norm. $|\mat{A}|$ is the determinant of matrix $\mat{A}$. $\mat{I}_N$ denotes the identity matrix of order $N$. 
$\delta(t)$ is the Dirac delta. 

\section{Transmission Model}
\label{section:transmission_model}

We assume that Alice and Bob extract a common key from observations of their shared channel $H$, assumed to be reciprocal. The channel $H$ is estimated based on the transmission of \textit{a priori} known pilots by Alice and Bob. We consider that the channel remains invariant during the transmission of each pilot symbol. Assuming a narrowband channel, the estimates of $H$ at Alice's and Bob's sides, respectively denoted by $X$ and $Y$, are given by
\begin{align*}
X&=H+W_X,\ Y=H+W_Y,
\end{align*}
where the additive noise samples $W_X$ and $W_Y$ are modeled as zero mean circularly-symmetric complex Gaussian (ZMCSCG) with variance $N_X$ and $N_Y$ respectively.  

The strategy of Eve consists in going as close as possible from Bob's antenna\footnote{Note that all of the following is symmetrical if Eve gets close to Alice instead of Bob.}. Then, Eve estimates her channel $H_Z$ between Alice's antenna and hers by intercepting the pilots sent from Alice to Bob. Since Eve is close to Bob, the channel from Alice to Eve will be spatially correlated with $H$ while the channel between Bob and Eve will experience a negligible correlation with $H$. Therefore, we neglect the pilot sent by Bob and received by Eve in the following as she cannot get any useful information from it. We define the channel estimate of Eve as 
\begin{align*}
Z=H_Z+W_Z,
\end{align*}
where $W_Z$ is modeled as ZMCSCG with variance $N_Z$. 
If Alice and Bob transmit a pilot of equal power and Alice, Bob and Eve use a similar receiver, one could expect a situation of equal noise variance $N_X=N_Y=N_Z$. On the other hand, Eve could use a more powerful receiver than Alice and/or Bob by having, \textit{e.g.}, a larger antenna size, a multi-antenna receiver or an amplifier with lower noise figure. This would result in a lower noise variance $N_Z$ and a higher SNR. Moreover, a different pilot power transmitted by Alice and Bob will induce variations in their noise variance.

We consider a memoryless source model for secret-key agreement \cite{Csiszar1993,bloch2011physical} as shown in Fig.~\ref{fig:source_model}. This implies that Alice, Bob and Eve observes $N$ independent and identically distributed (i.i.d.) repetitions of the random variables measurements $X$, $Y$ and $Z$, giving $X^N=(X_1,...,X_N)$, $Y^N=(Y_1,...,Y_N)$ and $Z^N=(Z_1,...,Z_N)$. Moreover, a noiseless authenticated public channel of unlimited capacity is available for communication. All parties have access to the public channel.

 \section{Channel Model}
\label{section:channel_model}

\begin{figure}[!t]  
	\centering
	
	\resizebox{0.9\textwidth}{!}{%
		{\includegraphics[clip, trim=0cm 12.5cm 8cm 0cm, scale=1]{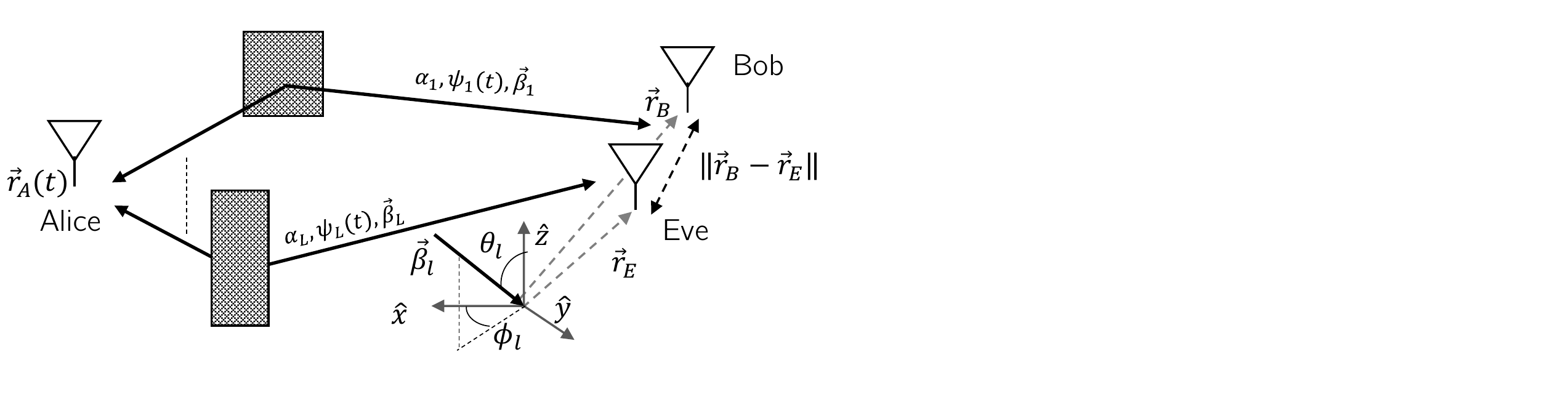}} 
	}
	\vspace{-1em}
	\caption{Non-LOS channel propagation model: Eve gets close to Bob to increase the spatial correlation between the channel from Alice to her antenna $H_Z(t)$ and the channel from Alice to Bob $H(t)$. Scatterers, Bob and Eve are static but Alice is moving.}
	\label{fig:channel_model}
	\vspace{-2em}
\end{figure}

In this section, we aim at providing a realistic model of the channel $H$ and $H_Z$ with particular emphasis on their correlation in time and space. The assumption of i.i.d. repeated measurements $X^N,Y^N,Z^N$ is well fulfilled in practice if the measurements of the channel are repeated in time with a sampling period that is large compared to the coherence time of the channel. This time can be related to the degree of mobility of Alice, Bob, Eve and the scattering environment. In this paper, we assume that Bob and Eve and scatterers are fixed while Alice is moving. This can model a typical situation where Bob is a base station and Alice is a user terminal. This implies that $H$ and $H_Z$ change over time, which leads to channel decorrelation in time. Alice is also assume to be in a non line-of-sight (LOS) situation.

As opposed to previous approaches \cite{wong2009secret,Ye2010,Chen2011,Jorswieck2013}, we do not consider a uniform distribution of scatterers in angle. Instead, we consider a model where the channels $H$ and $H_Z$ can be described by the combination of $L$ paths, as shown in Fig.~\ref{fig:channel_model}. The $l$-th path is characterized by an azimuth angle $\phi_{l}$ and elevation angle $\theta_{l}$ at Bob and Eve side. Bob and Eve belong to a local area, situated in the far field from scatterers, so that the angles are identical at Bob and Eve. Moreover they are assumed to remain constant in time over the $N$ measurements. The mobility of ALice induces a phase drift of each multipath component, common at Bob and Eve, that we denote by $\psi_l(t)$. Bob and Eve positions are denoted by $\vec{r}_{B} \in \mathbb{R}^{3\times 1}$ and $\vec{r}_{E} \in \mathbb{R}^{3\times 1}$ respectively in their local area coordinate system. Alice, Bob and Eve are each equipped with a single isotropic antenna. Under previous assumptions, the narrowband multipath channels $H$ and $H_Z$ at time $t$ can be modeled as \cite{molisch2012wireless}
\begin{align*}
H(t)&=\sum_{l=1}^L \alpha_l e^{\jmath \psi_l(t)} e^{-\jmath \vec{\beta}_l\cdot \vec{r}_B}\\
H_Z(t)&=\sum_{l=1}^L \alpha_l e^{\jmath \psi_l(t)} e^{-\jmath \vec{\beta}_l\cdot \vec{r}_E},
\end{align*}
where $\vec{\beta}_l \in \mathbb{R}^{3\times 1}$ is the wave vector associated to path $l$ at Bob/Eve side, which is directly related to the carrier wavelength $\lambda$ and points in direction $(\phi_l,\theta_l)$. $\alpha_l$ is the complex gain of path $l$.

In the following, in accordance with conventional approaches in the propagation literature \cite{durgin2003space}, we consider multipath components as stochastic. Moreover, we assume uncorrelated scatterers with a uniformly distributed phase so that $\mathbb{E}(\alpha_l)=0$ and $\mathbb{E}(\alpha_l \alpha_{l'}^*)=P_l\delta_{l-l'}$ with $P_l=\mathbb{E}(|\alpha_l|^2)$. We also define $P=\sum_l P_l$ as the average power of scattered paths. Non LOS measurement campaigns have shown that the channels $H(t)$ and $H_Z(t)$ can be accurately modeled with a zero-mean Gaussian distribution, especially for large values of $L$. Therefore, we assume that the random vector $(H(t), H_Z(t))^T$ follows a joint circularly symmetric Gaussian distribution with zero mean and covariance matrix given by
\begin{align*}
\mathbb{E}\begin{pmatrix}
H(t)\\ H_Z(t)
\end{pmatrix}= \begin{pmatrix}
0\\
0
\end{pmatrix} ,\ \mat{C}_{HH_Z}=P\begin{pmatrix}
1& \rho\\
\rho^*& 1
\end{pmatrix}.
\end{align*}
We now study the spatial correlation coefficient $\rho$ between $H(t)$ and $H_Z(t)$. Under previous assumptions, we can write
\begin{align}
\rho&= \mathbb{E}\left[H(t)H_Z^*(t)\right]
=\frac{1}{P} \sum_{l=1}^L P_l e^{-\jmath \vec{\beta}_l \cdot (\vec{r}_B-\vec{r}_E ) }\label{eq:rho}.
\end{align}
Defining the normalized angular power density function $f(\Omega)$ at Bob and Eve (per steradian) as
\begin{align*}
f(\Omega)&=\frac{1}{P} \sum_{l=1}^L P_l \delta\left(\Omega-\Omega_l\right)\\
\delta\left(\Omega-\Omega_l\right)&=\frac{1}{\sin \theta_l} \delta\left(\phi-\phi_l\right)\delta\left(\theta-\theta_l\right),
\end{align*}
we can rewrite (\ref{eq:rho}) as
\begin{align}
\rho= \int_{\Omega} f(\Omega) e^{-\jmath \vec{\beta}_l \cdot (\vec{r}_B-\vec{r}_E ) } d\Omega, \label{eq:def_rho2}
\end{align}
where the differential $d\Omega$ can be formulated in spherical coordinates as $d\Omega=\sin \theta d\theta d\phi$. Note that the wave vectors $\vec{\beta}_l$ depend on $\Omega$ through the angles $(\phi_l,\theta_l)$. Since $P=\sum_{l=1}^L P_l$, we have $\int_{\Omega} f(\Omega)d\Omega=1$ so that $0\leq |\rho|\leq 1$.

\begin{figure}[!t]
	\centering
	\resizebox{0.49\textwidth}{!}{%
		\Large
		\input{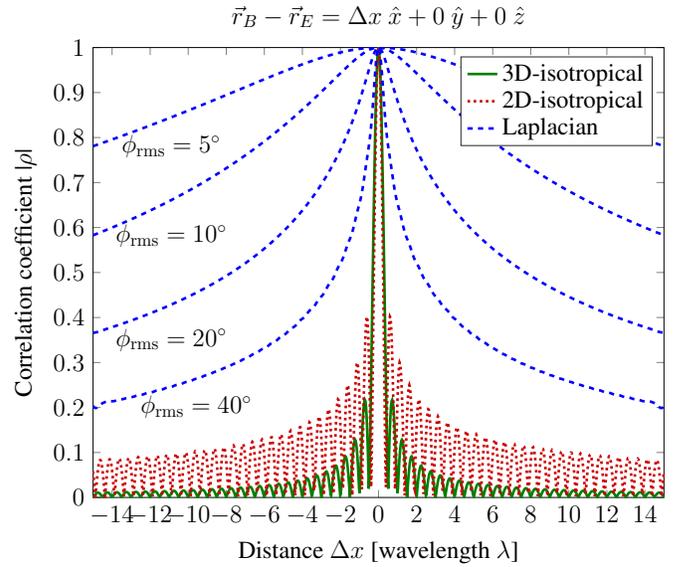}
	}
	\caption{Correlation coefficient $|\rho|$ as a function of the distance $\Delta x=\|\vec{r}_B-\vec{r}_E\|$ along the $\hat{x}$ axis. Three types of angular distributions are considered: (i) 3D-isotropic, $\rho= {\text{sinc}\left(\frac{2\pi \Delta x}{\lambda}  \right)}$, (ii) 2-D isotropic, $\rho=J_0\left(\frac{2\pi \Delta x}{\lambda}\right)$ and (iii) $\rho=(\ref{eq:def_rho2})$ with $f(\Omega)$ a Laplacian distribution with elevation spread $\theta_{\mathrm{rms}}=5^{\circ}$ and different angular spreads $\phi_{\mathrm{rms}}$.}
	\label{fig:rho_Laplacian}
	\vspace{-1em}
\end{figure}

The formulation (\ref{eq:def_rho2}) in terms of $f(\Omega)$ is convenient as it allows to represent both specular components and continuous spectrum, which can occur in the case of diffuse scattering. Typically, the value of $\rho$ will decrease as (i) the distance of Eve with respect to Bob $\|\vec{r}_B-\vec{r}_E\|$ increases and/or (ii) as the distribution $f(\Omega)$ gets uniform over $(\phi,\theta)$. The worst-case in terms of secrecy occurs in the extreme case $f(\Omega)=\delta(\Omega-\Omega_1)$ (only one incoming direction for the scattered paths) implying that $|\rho|=1$. The opposite extreme case is an isotropic distribution over azimuth and elevation (3D-isotropic), \textit{i.e.}, $f(\Omega)=\frac{1}{4\pi}$, which leads to the well-known result \cite[p. 49]{johnson1993array}
\begin{align}
\rho= {\text{sinc}\left(\frac{2\pi \|\vec{r}_B-\vec{r}_E\|}{\lambda}  \right)}, \label{eq:def_rho_iso}
\end{align}
where $\text{sinc}(x)=\sin x/x$. This shows that $\rho$ becomes negligible after a few $\lambda$. Similarly, if scattered paths are coming from a fixed horizontal elevation $\theta=\pi/2$ but are uniformly distributed in azimuth (2D-isotropic), \textit{i.e.}, $f(\Omega)=\frac{1}{2\pi}\delta(\theta-\pi/2)$, we get another well-known result \cite{stuber1996principles}
\begin{align*}
\rho=J_0\left(\frac{2\pi d}{\lambda}\right),
\end{align*}
where $J_0(.)$ is the zero-order Bessel function and $d$ is the distance between Eve and Bob in the horizontal plane. Here again, $\rho$ becomes negligible if $d$ is larger than a few $\lambda$. This was the justification of many works suggesting that Eve does not get any useful information about $H$ from $Z$ as soon as she is a few centimeters away from Bob for conventional radio-frequency bands \cite{AzimiSadjadi2007,wong2009secret,Ye2010,Chen2011,Jorswieck2013}. This implies that Eve can only learn about $H$ and thus the secret-key from the public discussion between Alice and Bob.

In this paper, we do not make this assumption, which can be too optimistic in terms of secrecy. In practice, the decay of $\rho$ as a function of $\|\vec{r}_B-\vec{r}_E\|$ will depend on the scattering environment, which is often far from being uniformly distributed in angle but rather clustered with specific angular spreads. As an example, we compare in Fig.~\ref{fig:rho_Laplacian} the decay of $\rho$ as a function of the distance between Eve and Bob, along the $\hat{x}$ axis ($\phi=0$ and $\theta=\pi/2$), and different angular distributions. In addition to the previously described 3D- and 2D-isotropic distributions, we also consider a more realistic Laplacian distribution in azimuth and elevation, centered in $(\phi,\theta)=(0,\pi/2)$, which is a common model for a base station \cite{Pedersen1997}
\begin{align*}
	f(\Omega)= \gamma e^{-\sqrt{2} \frac{|\phi|}{\phi_{\mathrm{rms}}}} \frac{1}{\sin \theta} e^{-\sqrt{2} \frac{|\theta-\pi/2|}{\theta_{\mathrm{rms}}}},
\end{align*}
where $\phi_{\mathrm{rms}}$ and $\theta_{\mathrm{rms}}$ are the azimuth and elevation angular spreads respectively and $\gamma$ is a normalization constant. According to recent 3GPP standard channel models \cite{3GPP_TR_38_901_v15}, typical values of $\phi_{\mathrm{rms}}$ range around 40$^{\circ}$ for an indoor office, in $[1^{\circ},10^{\circ}]$ for a rural environment and in $[10^{\circ},40^{\circ}]$ for an urban micro/macro cell environment while a typical value for the elevation angular spread is $\theta_{\mathrm{rms}}=5^{\circ}$. We can clearly see in Fig.~\ref{fig:rho_Laplacian} that the 3D- and 2D-isotropic distributions underestimate the spatial correlation between Bob and Eve. In other words, considering these models overestimates the secret-key capacity in scenarios of practical relevance. For a typical cellular carrier frequency of 1 GHz, $\lambda=30$ cm and Eve could be placed at $~10 \lambda=3$ m while still having a significant correlation with $H$. Moreover, as explained earlier, Eve could also use a more powerful receiver than Alice and Bob resulting in a lower noise variance $N_Z$.

\section{Secret-Key Capacity}
\label{section:secret_key_capacity}

The secret-key capacity $S(X;Y||Z)$ is defined as the maximum rate at which Alice and Bob can agree on a secret-key while keeping the rate at which Eve obtains information about the key arbitrarily small for sufficiently large $N$. Moreover, Alice and Bob should agree on a common key with high probability and the key should approach the uniform distribution. We refer to \cite{Csiszar1993,Maurer1993,bloch2011physical} for a formal definition. 

As explained above, we consider that Eve gets useful information from her observation $Z$ over $H$. This implies that the secret-key capacity is not simply equal to $I(X;Y)$, as opposed to many previous works. Finding the general expression of the secret-key capacity for a given distribution of $X,Y,Z$ is still an open problem. From \cite{Csiszar1993,Maurer1993} \cite[Prop. 5.4]{bloch2011physical}, the secret-key capacity, expressed in the number of generated secret bits per channel observation, can be lower and upper bounded as follows
\begin{align}
S(X;Y||Z)&\geq  I(X;Y)-\min\left[I(X;Z), I(Y;Z) \right] \label{eq:lower_bounds}\\
S(X;Y||Z)&\leq \min\left[ I(X;Y),I(X;Y|Z) \right].  \label{eq:upper_bounds}
\end{align}
The lower bound (\ref{eq:lower_bounds}) implies that if Eve has less information about $Y$ than Alice or respectively about $X$ than Bob, such a difference can be leveraged for secrecy \cite{Maurer1993}. Moreover, this rate can be achieved with one-way communication. On the other hand, the upper bound (\ref{eq:upper_bounds}) implies that the secret-key rate cannot exceed the mutual information between Alice and Bob. Moreover, the secret-key rate cannot be higher than the mutual information between Alice and Bob if they happened to learn Eve's observation $Z$. 

In particular cases, the lower and upper bounds can become tight \cite{Csiszar1993,Maurer1993,bloch2011physical}. In the next subsections, we evaluate the lower and upper bounds of (\ref{eq:lower_bounds}) and (\ref{eq:upper_bounds}), and their simplification in the cases where the bounds become tight. To do this, we use the fact that, from the system model detailed in previous sections, the random variables $X$, $Y$ and $Z$ are jointly circularly symmetric Gaussian distributed. 
This implies that the entropy of these random variables only depend on their covariance, which is equivalent to their correlation given their zero mean. The entropy of a circularly symmetric Gaussian with covariance $\mat{C}$ is $\log_2( |\pi e\mat{C}|)$, where $e$ is the Euler number. 

\subsection{Lower Bound}

The random variables $X$ and $Y$ are jointly Gaussian distributed with covariance
\begin{align*}
\mat{C}_{XY}=\begin{pmatrix}
P+N_X& P\\
P& P+N_Y
\end{pmatrix}.
\end{align*}
From this distribution, we find
\begin{align}
I(X;Y)&=H(X)+H(Y)-H(XY)\label{eq:I_XY}\\
&=\log_2\left(1+\frac{P^2}{(P+N_X)(P+N_Y)-P^2}\right)
.\nonumber
\end{align}
Moreover, $X$ and $Z$ are jointly Gaussian with covariance
\begin{align*}
\mat{C}_{XZ}=\begin{pmatrix}
P+N_X& \rho P\\
\rho^*P& P+N_Z
\end{pmatrix}.
\end{align*}
This leads to the mutual information
\begin{align*}
I(X;Z)
&=\log_2\left(1+\frac{|\rho P|^2}{(P+N_X)(P+N_Z)-|\rho P|^2}\right).
\end{align*}
Using a similar methodology for $Y$, we find
\begin{align*}
I(Y;Z)&=\log_2\left(1+\frac{|\rho P|^2}{(P+N_Y)(P+N_Z)-|\rho P|^2}\right).
\end{align*}
In the end, we find that the lower bound in (\ref{eq:lower_bounds}) is equal to
\begin{align*}
S(X;Y||Z)&\geq \log_2\left(\frac{1+\frac{P^2}{(P+N_X)(P+N_Y)-P^2}}{1+\frac{|\rho P|^2}{(P+\max(N_X,N_Y))(P+N_Z)-|\rho P|^2}}\right).
\end{align*}
As soon as $|\rho|<1$, $S(X;Y||Z)$ is unbounded and goes to infinity as $P\rightarrow +\infty$. Indeed, as $P\rightarrow +\infty$, $I(X;Y)\rightarrow +\infty$ while $I(X;Z)$ and $I(Y;Z)$ converge to $\log_2\left(1+\frac{|\rho|^2}{1-|\rho|^2}\right)$, which is bounded for $|\rho|<1$. Note that the lower bound is not restricted to be positive (as will be shown in Section~\ref{section:Numerical_validation}), in which case it becomes useless. We can find the condition on the minimum noise variance at Eve $N_Z$ for having a larger-than-zero lower bound
\begin{align}
N_Z&> P(|\rho |^2-1) +|\rho |^2\min(N_X,N_Y). \label{eq:lower_bound_N_Z}
\end{align}
In the worst-case, $|\rho|=1$ and $N_Z$ has to be larger than the minimum of the noise variances of Alice and Bob. We can invert (\ref{eq:lower_bound_N_Z}) to find the maximal correlation coefficient $|\rho|^2$ to have a larger-than-zero lower bound
\begin{align*}
|\rho |^2&< \frac{P+N_Z}{P+\min(N_X,N_Y)}.
\end{align*}
From the definition of $\rho$ in (\ref{eq:def_rho2}) and for a given propagation environment inducing a specific angular power distribution $f(\Omega)$, the last equation can be related to the minimal admissible distance $\|\vec{r}_B-\vec{r}_E\|$ between Eve and Bob.

\subsection{Upper Bound}

To evaluate the upper bound in (\ref{eq:upper_bounds}), we only need to evaluate $I(X;Y|Z)$ as we already evaluated the expression $I(X;Y)$. To do this, first note that $X$, $Y$ and $Z$ are jointly Gaussian distributed with covariance matrix
\begin{align*}
\mat{C}_{XYZ}=\begin{pmatrix}
P+N_X& P & \rho P\\
P& P+N_Y & \rho P\\
\rho^* P& \rho^* P& P+N_Z
\end{pmatrix},
\end{align*}
which gives
\begin{align*}
I(X;Y|Z)
&=H(XZ)+H(YZ)-H(Z)-H(XYZ)\\
&=\log_2\left(\frac{|\mat{C}_{XZ}||\mat{C}_{YZ}|}{(P+N_Z)|\mat{C}_{XYZ}|}\right).
\end{align*}
The upper bound is then given by the minimum of $I(X;Y|Z)$ and $I(X;Y)$. 
It is possible to prove that the condition $I(X;Y|Z)\leq I(X;Y)$ is always verified under the assumptions of our channel model\footnote{Proof is omitted due to space constraints.} and thus
\begin{align*}
S(X;Y||Z)\leq \log_2\left(\frac{|\mat{C}_{XZ}||\mat{C}_{YZ}|}{(P+N_Z)|\mat{C}_{XYZ}|}\right).
\end{align*}
One should note that the situation $I(X;Y|Z)> I(X;Y)$ could be possible in another context. This would imply that the knowledge of Eve's observation could help Alice and Bob to generate secrecy. We refer to \cite{Yeung1991} for an analysis of the quantity $I(X;Y|Z)-I(X;Y)$ and to \cite{Csiszar1993} for its implication in terms of secrecy.

\subsection{Tight Bound}

In our context, three particular cases can be distinguished.

1) $\rho=0$, Eve does not learn anything about $H$ from $Z$, which becomes independent from $X$ and $Y$. This leads to the trivial result $S(X;Y||Z)=I(X;Y)$, given in (\ref{eq:I_XY}) and as considered in \cite{wong2009secret,Ye2010,Chen2011,Jorswieck2013}.

2) $N_Y=0,\ Y=H$, this implies that $X\rightarrow Y \rightarrow Z$ forms a Markov chain, which leads to $I(X;Y|Z)=I(X;Y)-I(X;Z)$ \cite[Corol. 4.1]{bloch2011physical} and
\begin{align*}
	S(X;Y||Z)&= \log_2\left(\frac{1+\frac{P}{N_X}}{1+\frac{|\rho P|^2}{(P+N_X)(P+N_Z)-|\rho P|^2}}\right).
\end{align*}
	
3) {$N_X=0,\ X=H$}, symmetrically as in 2), we find
\begin{align*}
	S(X;Y||Z)&= \log_2\left(\frac{1+\frac{P}{N_Y}}{1+\frac{|\rho P|^2}{(P+N_Y)(P+N_Z)-|\rho P|^2}}\right).
\end{align*}
Cases 2) and 3) let us expect that the bounds become tight as the receiver of Alice or Bob is significantly noisier than the one of the other.

\section{Numerical Validation}
\label{section:Numerical_validation}

\begin{figure}[!t]
	\centering
	\resizebox{0.49\textwidth}{!}{%
		\Large
		\input{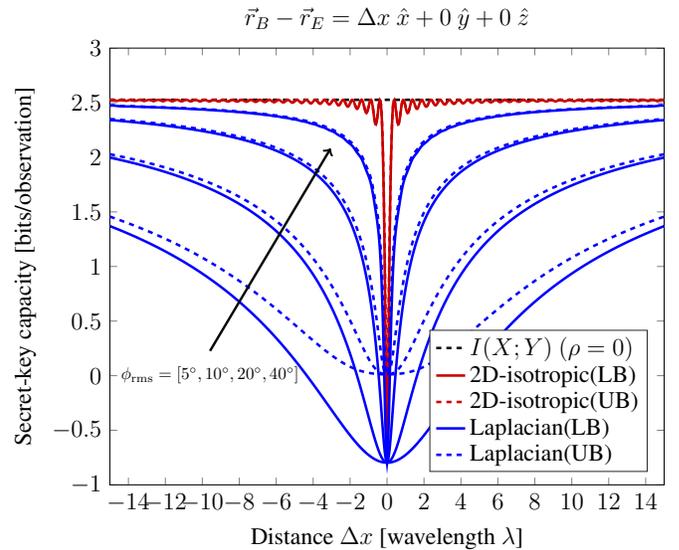}
	}
	\caption{Lower bound (LB) and upper bound (UB) for secret-key capacity as a function of propagation environments considered in Fig.~\ref{fig:rho_Laplacian}. The curve $I(X,Y)$ corresponds to the case of independent observations at Eve ($\rho=0$).}
	\label{fig:secret_key_capacity}
	\vspace{-1em}
\end{figure}

We evaluate numerically the secret-key capacity in Fig.~\ref{fig:secret_key_capacity} based on the formulas derived in Section~\ref{section:secret_key_capacity} for the lower/upper bounds (LB/UB) and relying on the channel models derived in Section~\ref{section:channel_model}. We consider the same angular distributions as in Fig.~\ref{fig:rho_Laplacian}. We recall that an angular Laplacian distribution is an accurate model for a base station with varying angular spreads as a function of the propagation environment. We consider signal-to-noise ratios $P/N_X=P/N_Y= 10$ dB at Alice and Bob while Eve is allowed to have a more power receiver that achieves $P/N_Z=20$ dB. For a given angular spread, Fig.~\ref{fig:secret_key_capacity} gives the admissible distance between Bob and Eve to achieve a given secret-key capacity.

Here again, we see that the general assumption of considering that Eve's observations $Z$ are independent of $X$ and $Y$ ($\rho =0$) is well verified if scatterers are 2D-isotropic distributed and if Eve is more than a wavelength away from Bob. However, for practical angular spreads at the base station, this assumption is typically not valid and too optimistic. This implies that lower secret-key rates are achievable in practice and privacy amplification should compensate for the information that Eves learns about the key not only from public discussion but also from her observations.


\section{Conclusion} \label{section:conclusion}

In this paper, we have studied the secret-key capacity based on the principle of channel reciprocity. We have shown that the assumption of full decorrelation of Eve's observations with respect to Alice and Bob is not always verified and critically depends on the propagation environment. Our simulation results show that, for practical propagation environments, the correlation of Eve's observations is non negligible implying a potentially significant reduction of the secret-key capacity.

\footnotesize
\bibliographystyle{IEEEtran}
\bibliography{IEEEabrv,IEEEreferences}

\begin{thebibliography}{10}
\providecommand{\url}[1]{#1}
\csname url@samestyle\endcsname
\providecommand{\newblock}{\relax}
\providecommand{\bibinfo}[2]{#2}
\providecommand{\BIBentrySTDinterwordspacing}{\spaceskip=0pt\relax}
\providecommand{\BIBentryALTinterwordstretchfactor}{4}
\providecommand{\BIBentryALTinterwordspacing}{\spaceskip=\fontdimen2\font plus
\BIBentryALTinterwordstretchfactor\fontdimen3\font minus
  \fontdimen4\font\relax}
\providecommand{\BIBforeignlanguage}[2]{{%
\expandafter\ifx\csname l@#1\endcsname\relax
\typeout{** WARNING: IEEEtran.bst: No hyphenation pattern has been}%
\typeout{** loaded for the language `#1'. Using the pattern for}%
\typeout{** the default language instead.}%
\else
\language=\csname l@#1\endcsname
\fi
#2}}
\providecommand{\BIBdecl}{\relax}
\BIBdecl

\bibitem{Wyner1975}
A.~D. {Wyner}, ``{The wire-tap channel},'' \emph{The Bell System Technical
  Journal}, vol.~54, no.~8, pp. 1355--1387, Oct 1975.

\bibitem{Csiszar1978}
I.~{Csiszar} and J.~{Korner}, ``Broadcast channels with confidential
  messages,'' \emph{IEEE Transactions on Information Theory}, vol.~24, no.~3,
  pp. 339--348, May 1978.

\bibitem{bloch2011physical}
M.~Bloch and J.~Barros, \emph{{Physical-layer security: from information theory
  to security engineering}}.\hskip 1em plus 0.5em minus 0.4em\relax Cambridge
  University Press, 2011.

\bibitem{Maurer1993}
U.~M. {Maurer}, ``{Secret key agreement by public discussion from common
  information},'' \emph{IEEE Transactions on Information Theory}, vol.~39,
  no.~3, pp. 733--742, May 1993.

\bibitem{Csiszar1993}
R.~{Ahlswede} and I.~{Csiszar}, ``{Common randomness in information theory and
  cryptography. I. Secret sharing},'' \emph{IEEE Transactions on Information
  Theory}, vol.~39, no.~4, pp. 1121--1132, July 1993.

\bibitem{AzimiSadjadi2007}
\BIBentryALTinterwordspacing
B.~Azimi-Sadjadi, A.~Kiayias, A.~Mercado, and B.~Yener, ``{Robust Key
  Generation from Signal Envelopes in Wireless Networks},'' in
  \emph{Proceedings of the 14th ACM Conference on Computer and Communications
  Security}, ser. CCS '07.\hskip 1em plus 0.5em minus 0.4em\relax New York, NY,
  USA: ACM, 2007, pp. 401--410. [Online]. Available:
  \url{http://doi.acm.org/10.1145/1315245.1315295}
\BIBentrySTDinterwordspacing

\bibitem{wong2009secret}
T.~F. Wong, M.~Bloch, and J.~M. Shea, ``{Secret sharing over fast-fading MIMO
  wiretap channels},'' \emph{EURASIP Journal on Wireless Communications and
  Networking}, vol. 2009, no.~1, p. 506973, 2009.

\bibitem{Ye2010}
C.~{Ye}, S.~{Mathur}, A.~{Reznik}, Y.~{Shah}, W.~{Trappe}, and N.~B.
  {Mandayam}, ``{Information-Theoretically Secret Key Generation for Fading
  Wireless Channels},'' \emph{IEEE Transactions on Information Forensics and
  Security}, vol.~5, no.~2, pp. 240--254, June 2010.

\bibitem{Chen2011}
C.~{Chen} and M.~A. {Jensen}, ``{Secret Key Establishment Using Temporally and
  Spatially Correlated Wireless Channel Coefficients},'' \emph{IEEE
  Transactions on Mobile Computing}, vol.~10, no.~2, pp. 205--215, Feb 2011.

\bibitem{Jorswieck2013}
E.~A. {Jorswieck}, A.~{Wolf}, and S.~{Engelmann}, ``{Secret key generation from
  reciprocal spatially correlated MIMO channels},'' in \emph{2013 IEEE Globecom
  Workshops (GC Wkshps)}, Dec 2013, pp. 1245--1250.

\bibitem{Mathur2008}
\BIBentryALTinterwordspacing
S.~Mathur, W.~Trappe, N.~Mandayam, C.~Ye, and A.~Reznik, ``{Radio-telepathy:
  Extracting a Secret Key from an Unauthenticated Wireless Channel},'' in
  \emph{Proceedings of the 14th ACM International Conference on Mobile
  Computing and Networking}, ser. MobiCom '08.\hskip 1em plus 0.5em minus
  0.4em\relax New York, NY, USA: ACM, 2008, pp. 128--139. [Online]. Available:
  \url{http://doi.acm.org/10.1145/1409944.1409960}
\BIBentrySTDinterwordspacing

\bibitem{Madiseh2009}
M.~{Ghoreishi Madiseh}, S.~{He}, M.~L. {Mcguire}, S.~W. {Neville}, and
  X.~{Dong}, ``{Verification of Secret Key Generation from UWB Channel
  Observations},'' in \emph{2009 IEEE International Conference on
  Communications}, June 2009, pp. 1--5.

\bibitem{Zhang2016}
J.~{Zhang}, R.~{Woods}, T.~Q. {Duong}, A.~{Marshall}, Y.~{Ding}, Y.~{Huang},
  and Q.~{Xu}, ``{Experimental Study on Key Generation for Physical Layer
  Security in Wireless Communications},'' \emph{IEEE Access}, vol.~4, pp.
  4464--4477, 2016.

\bibitem{Chou2010}
T.~{Chou}, S.~C. {Draper}, and A.~M. {Sayeed}, ``{Impact of channel sparsity
  and correlated eavesdropping on secret key generation from multipath channel
  randomness},'' in \emph{2010 IEEE International Symposium on Information
  Theory}, June 2010, pp. 2518--2522.

\bibitem{Zhang2017}
J.~{Zhang}, B.~{He}, T.~Q. {Duong}, and R.~{Woods}, ``{On the Key Generation
  From Correlated Wireless Channels},'' \emph{IEEE Communications Letters},
  vol.~21, no.~4, pp. 961--964, April 2017.

\bibitem{3GPP_TR_38_901_v15}
``{3GPP TR 38.901 v15.0.0},'' Tech. Rep., 2018.

\bibitem{Thoen2002}
S.~{Thoen}, L.~{Van der Perre}, and M.~{Engels}, ``{Modeling the channel
  time-variance for fixed wireless communications},'' \emph{IEEE Communications
  Letters}, vol.~6, no.~8, pp. 331--333, Aug 2002.

\bibitem{molisch2012wireless}
A.~F. Molisch, \emph{{Wireless communications}}.\hskip 1em plus 0.5em minus
  0.4em\relax John Wiley \& Sons, 2012, vol.~34.

\bibitem{durgin2003space}
G.~D. Durgin, \emph{{Space-time wireless channels}}.\hskip 1em plus 0.5em minus
  0.4em\relax Prentice Hall Professional, 2003.

\bibitem{johnson1993array}
D.~H. Johnson and D.~E. Dudgeon, \emph{{Array signal processing: concepts and
  techniques}}.

\bibitem{stuber1996principles}
G.~L. St{\"u}ber and G.~L. St{\`e}uber, \emph{{Principles of mobile
  communication}}.\hskip 1em plus 0.5em minus 0.4em\relax Springer, 1996,
  vol.~2.

\bibitem{Pedersen1997}
K.~I. {Pedersen}, P.~E. {Mogensen}, and B.~H. {Fleury}, ``{Power azimuth
  spectrum in outdoor environments},'' \emph{Electronics Letters}, vol.~33,
  no.~18, pp. 1583--1584, Aug 1997.

\bibitem{Yeung1991}
R.~W. {Yeung}, ``{A new outlook on Shannon's information measures},''
  \emph{IEEE Transactions on Information Theory}, vol.~37, no.~3, pp. 466--474,
  May 1991.

\end{thebibliography}

\end{document}